\documentclass[preprint,superscriptaddress, nofootinbib, aps,pra]{revtex4}
\usepackage{braket}             
\usepackage{graphicx}           
\usepackage{bm}                 
\usepackage{amsmath,amssymb,amsthm}
\usepackage[usenames, dvipsnames]{color}
\usepackage{xcolor}
\usepackage{hyperref}           

\usepackage{mathtools}
\usepackage{bigstrut,multirow,rotating}
\usepackage{soul}               
\usepackage{mathrsfs}
\usepackage{subfigure}
\usepackage{ulem}
\DeclareMathOperator{\tr}{tr}   
\newcommand{\nn}{\nonumber}     
\usepackage{colonequals} 
\usepackage{graphics,epsfig}

\begin{document}
\title{Entanglement harvesting of circularly accelerated detectors with a reflecting boundary}
\author{Runhu Li$^{1}$, Zixu Zhao\footnote{Corresponding author. zhao$_{-}$zixu@yeah.net}}
\affiliation{School of Science, Xi'an University of Posts and Telecommunications, Xi'an 710121, China}

\begin{abstract}

We study the properties of the transition probability for a circularly accelerated detector that interacts with the massless scalar fields in the presence of a reflecting boundary. As the trajectory radius increases, the transition probability may exhibit some peaks under specific conditions, which can lead to the possibility of the identical result for different trajectory radius with the same acceleration and energy gap. These behaviors can be characterized by certain critical values. Furthermore, we analyze the entanglement harvesting phenomenon for two circularly accelerated detectors with a boundary. We consider that the two detectors are rotating around a common axis with the same acceleration, trajectory radius and angular velocity. When the detectors are close to the boundary, there may exist two peaks in entanglement harvesting. Interestingly, as trajectory radius increases, entanglement harvesting in some situations first decreases to zero, then remains zero, and finally increases to a stable value. Our results show that the features observed properties of the detectors are closely related to the vacuum fluctuations of the field and states of the detectors. Under appropriate conditions, circular motion can be used to simulate the results of uniform acceleration.

\end{abstract}
\pacs{03.67.Bg,04.62.+v,03.65.Ud,03.67.-a,11.10.-z}
\maketitle
\section{Introduction}
	
In 1932, von Neumann completed the mathematical foundations of non-relativistic quantum description~\cite{Neumann:1932}. In 1935, Einstein, Podolsky, and Rosen (EPR) attempted to show that ``the wave function does not provide a complete description of the physical reality"~\cite{EPR:1935}. The EPR paper inspired extensive discussion about the foundations of quantum mechanics. Following the EPR paper, Schr\"{o}dinger introduced the term ``Verschr\"{a}nkung"~\cite{Schrodinger:1935}, and discussed the concept of ``entanglement''~\cite{Schrodinger:1935English}. Wu and Shaknov studied the angular correlation of scattered annihilation radiation~\cite{Wu:1950}. A variant of the EPR thought experiment was later proposed~\cite{Bohm:1951,Bohm:1952I,Bohm:1952II,Bohm&Aharonov:1957}. In 1964, Bell's inequality demonstrated that quantum physics predicts correlations that violate this inequality~\cite{Bell:1964}. Freedman and Clauser performed the first rudimentary experiment designed to test Bell's theorem in 1972, which was only a limited test~\cite{Freedman&Clauser:1972}. For example, the experiment used polarizers that were preset. Aspect and his collaborators performed the Bell test that removed this limitation~\cite{Aspect&Grangier&Roger:1982,Aspect&Dalibard&Roger:1982}. Zeilinger's team achieved the first realization of quantum teleportation of an independent qubit~\cite{Zeilinger:1997}.

In recent years, it has been realized that the vacuum can be a resource of entanglement. Summers and Werner found that the vacuum state in free quantum field theory violates Bell's inequalities maximally~\cite{Summers:1985,Summers:1987I,Summers:1987II}. Valentini showed that a pair of initially uncorrelated atoms, separated by a distance $R$, develop non-local statistical correlations in a time $t<R/c$, which can be understood in terms of non-locally-correlated vacuum field fluctuations~\cite{Valentini:1991}. Reznik showed that entanglement persists between disconnected regions in the vacuum, and vacuum entanglement becomes a physical operational quantity, by transforming vacuum entanglement to pairs of probes or atom-like system~\cite{Reznik:2003}. Reznik, Retzker and Silman presented a physical effect of vacuum fluctuations which is associated with quantum nonlocality~\cite{Reznik:2005}. Entanglement can be used to detect spacetime curvature, and quantum fields in the Minkowski vacuum are entangled with respect to local field modes~\cite{VM:2009}. Entanglement between the future and the past in the quantum vacuum has been studied~\cite{Olson and Ralph:2011}. Mart\'{i}n-Mart\'{i}nez and Menicucci considered the extraction of entanglement from a quantum field by coupling to local detectors and how this procedure can be used to distinguish curvature from heating by their entanglement signature~\cite{MM:2012}. Hu, Lin, and Louko studied relativistic quantum information in detectors-field interactions~\cite{HLL:2012}. Mart\'{i}n-Mart\'{i}nez, Brown, Donnelly, and Kempf proposed a protocol by which entanglement can be extracted repeatedly from a quantum field, and they called this protocol entanglement farming in analogy with prior work on entanglement harvesting~\cite{MBDK:2013}. Salton, Mann, and Menicucci investigated entanglement harvested from a quantum field through local interaction with Unruh-DeWitt (UDW) detectors undergoing linear acceleration~\cite{Salton-Man:2015}. Pozas-Kerstjens and Mart\'{i}n-Mart\'{i}nez analyzed the harvesting of entanglement and classical correlations from the quantum vacuum to particle detectors~\cite{KM:2015}.

Entanglement harvesting has been studied extensively~\cite{MMST:2016,MMS:2016,Zhjl:2018,Zhjl:2019,CW:2019,KMK:2019,ZJL:2020,CW:2020,LZH:2021,Bozanic:2023}. Zhang and Yu studied entanglement harvesting for detectors in circular motion~\cite{ZJL:2020}. Modifications to vacuum fluctuations may lead to rich features in the observed properties of the detectors. In this paper, we study the entanglement harvesting phenomenon for two circularly accelerated detectors in the presence of a reflecting boundary.  The particle detector interacting with vacuum quantum fields can be described using the well-known Unruh-DeWitt model~\cite{DeWitt:1979}. We aim to further explore the role of these parameters in detail.

The plan of the work is the following. In Sec. II, we review the basic formulae for UDW detectors locally interacting with vacuum scalar fields. In Sec. III, we consider the transition probabilities of circularly accelerated UDW detectors with a reflecting boundary. In Sec. IV, we study the entanglement harvesting phenomenon for a pair of coaxial accelerating detectors along circular trajectories with a reflecting boundary. We conclude in the last section with our main results. For convenience, we employ the natural units $\hbar = c = 1$ .

\section{The basic formulas}

In this section, we will consider a pair of point-like two-level atoms (labeled by $A$ and $B$) interacting locally with a quantum scalar filed. The two-level atom with the ground state $\ket{0}_D$ and excited state $\ket{1}_D$ separated by an energy gap $\Omega_D$ can be modeled with the UDW detector, where the subscript $D$ specifies which UDW detector we are considering. The spacetime trajectory $x_D(\tau_D)$ of the detector is parametrized in terms of its proper time. Then the interacting Hamiltonian for such a detector locally coupling with a massless scalar field $\phi\big[x_D(\tau_D)\big]$ has the following form in the interaction picture
\begin{equation} \label{Int2}
	 H_D(\tau_D)=\lambda\chi_D(\tau_D)\Big(e^{-i\Omega_D\tau_D}\sigma^-+e^{i\Omega_D\tau_D}\sigma^+\Big)\otimes\phi\big[x_D(\tau_D)\big]\;,~~D\in\{A,B\}
\end{equation}
where $\lambda$ is the coupling strength and $\chi_D(\tau_D):=e^{-\tau_D^2/(2\sigma_D^2)}$ is a Gaussian switching function which controls the duration of interaction via parameter $\sigma_D$. Particularly, $\sigma^-=\ket{0}_D\bra{1}_D$ and $\sigma^+=\ket{1}_D\bra{0}_D$ denote the ladder operators acting on the Hilbert space of the detector.

We assume that detectors $A$ and $B$ in their ground states and the scalar field in vacuum state $\ket{0}$ before the interaction begins. Therefore the initial joint state of the detectors and the field can be written as $\ket{\Psi}=\ket{0}_A\ket{0}_B\ket{0}$. We assume that the two detectors have completely identical energy gap $\Omega=\Omega_A=\Omega_B$ and switching parameter $\sigma=\sigma_A=\sigma_B$ for simplicity. According to the detector-field interaction Hamiltonian Eq.~(\ref{Int2}), the composite system (two detectors plus the field) will undergo the unitary evolution, where the evolution operator satisfies
\begin{equation}\label{U}
	U:={\cal{T}} \exp\Big[-i\int{dt}\Big(\frac{d\tau_A}{dt}H_A(\tau_A)+
	\frac{d\tau_B}{dt}{H_B}(\tau_B)\Big)\Big]\;,
\end{equation}
here ${\cal{T}}$ represents the time ordering operator. Based on the perturbation theory, in the basis $\{\ket{0}_A\ket{0}_B,\ket{0}_A\ket{1}_B,\ket{1}_A\ket{0}_B,\ket{1}_A\ket{1}_B\}$, the final reduced density matrix of the system (two detectors) can be obtained by tracing out the field degrees of freedom, after some manipulations~\cite{MMST:2016,Zhjl:2018,Zhjl:2019}
\begin{align}\label{rhoAB}
	\rho_{AB}:&=\tr_{\phi}\big(U\ket{\Psi}\bra{\Psi}U^{\dagger}\big)\nonumber\\
	&=\begin{pmatrix}
		1-P_A-P_B & 0 & 0 & X \\
		0 & P_B & C & 0 \\
		0 & C^* & P_A & 0 \\
		X^* & 0 & 0 & 0 \\
	\end{pmatrix}+{\mathcal{O}}(\lambda^4)\;.
\end{align}
In the reduced density matrix $\rho_{AB}$, the transition probability $P_D$ reads
\begin{equation}\label{probty}
	P_D:=\lambda^2\iint{d\tau_D}{d\tau_D'}\chi_D(\tau_D)\chi_D(\tau_D')e^{-i\Omega(\tau_D-\tau_D')}W(x_D(\tau_D),x_D(\tau_D'))\;\;\;\;
	D\in\{A,B\}\;,
\end{equation}
and the correlation terms $C$ and $X$ read
\begin{align}
	C:=&\lambda^2 \iint dt  dt' \,  \frac{\partial\tau_B}{\partial{t}} \frac{\partial\tau_A}{\partial{t'}} \chi_B(\tau_B(t))  \chi_A(\tau_A(t')) e^{i \left[ \Omega\tau_B(t)-\Omega \tau_A(t'\right)]} W\!\left(x_A(t') , x_B(t)\right)\;,
\end{align}

\begin{align}\label{defX}
	X&\colonequals-\lambda^2  \iint_{t>t'} dt  dt'  \bigg[
	\frac{\partial\tau_B}{\partial{t}} \frac{\partial\tau_A}{\partial{t'}} \chi_B(\tau_B(t)) \chi_A(\tau_A(t'))e^{-i\left[\Omega\tau_B(t)+\Omega\tau_A(t')\right]}W\!\left(x_A(t'), x_B(t)\right) \nn \\
	& \quad \qquad \qquad \qquad \quad  + \frac{\partial\tau_A}{\partial{t}}\frac{\partial\tau_B}{\partial{t'}} \chi_A(\tau_A(t)) \chi_B(\tau_B(t'))e^{-i\left[\Omega \tau_A(t)+\Omega\tau_B(t')\right]} W\!\left(x_B(t'),x_A(t)\right)\bigg],
\end{align}
where $W(x,x'):=\bra{0}\phi(x)\phi(x')\ket{0}$ denotes the Wightman function of the field.

Based on the entanglement harvesting protocol~\cite{Salton-Man:2015}, we employ the concurrence as a measure of entanglement~\cite{Wootters:1998}, which can quantify the entanglement harvested by the detectors via local interaction with the fields. For the $X$-like density matrix given in Eq.~(\ref{rhoAB}), the concurrence takes a form~\cite{MMST:2016}
\begin{equation}\label{con1}
	\mathcal{C}(\rho_{AB})=2\max\big\{0,|X|-\sqrt{P_AP_B}\big\}+{\mathcal{O}}(\lambda^4)\;.
\end{equation}
The concurrence $\mathcal{C}(\rho_{AB})$ depends on the non-local correlations $X$ and the transition probabilities $P_A$ and $P_B$, which is determined by the Wightman function of scalar fields. We will investigate the entanglement harvesting phenomenon for a pair of accelerated detectors moving in a circle near a boundary.

\section{The transition probabilities of circularly accelerated UDW detectors near the reflecting boundary}

\begin{figure*}[!ht]
\centering
\includegraphics[width=0.4\textwidth]{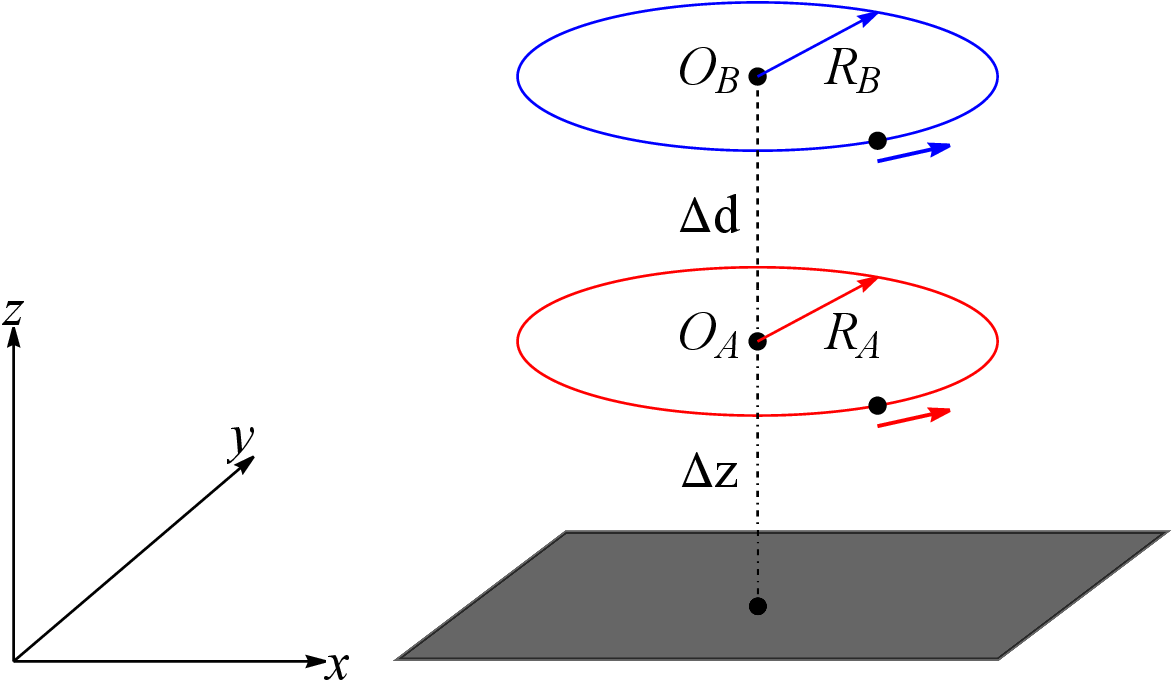}
\caption{The circular trajectories of two UDW detectors $A$ and $B$ are considered in flat spacetime. Here, the boundary lies in the $xy$-plane, and the two detectors share a common rotational axis.}\label{orbit}
\end{figure*}

We present the circular trajectories of the two detectors in Fig.~\ref{orbit}. Now, we first analyze the transition probability of a detector. For the convenience of discussion, we suppose that the plane boundary is located at $z=0$, and the accelerated UDW detector performs circular motion parallel to the $xy$-plane with a distance $\Delta z$ away from the boundary. The spacetime trajectory describing such detector parameterized by the proper time $\tau_D$ is
\begin{equation}\label{traj1}
x_D:=\{t=\gamma_D\tau_D\;,~~x=R_D\cos(\omega_D\gamma_D\tau_D)\;,~~y=R_D\sin(\omega_D\gamma_D\tau_D)\;,~~z={\Delta z}\}\;,
\end{equation}
where $R_D$ is the radius of the circular trajectory and $\omega_D$ represents the angular velocity of the detector moving in a circle, and $\gamma_D=1/\sqrt{1-R_D^2\omega_D^2}$ denotes the Lorentz factor. In the detector's frame, the magnitude of acceleration is $a_D=\gamma_D^2\omega_D^2R_D=\gamma_D^2v_D^2/R_D$, with the magnitude of linear velocity satisfies $v_D=|\omega_D|R_D<1$. It should be noted that $R_D$, $\omega_D$, $a_D$ and $v_D$ are not completely independent.

In a four dimensional Minkowski spacetime, the Wightman function for massless scalar fields with a reflecting boundary can be given by~\cite{Birrell:1982}
\begin{equation}\label{wightman1}
\begin{aligned}
W(x,x')=&-\frac{1}{4\pi^2}\big[\frac{1}{(t-t'-i\epsilon)^2-(x-x')^2-(y-y')^2-(z-z')^2}\\
&-\frac{1}{(t-t'-i\epsilon)^2-(x-x')^2-(y-y')^2-(z+z')^2}\big]\;.
\end{aligned}
\end{equation}

Substituting the trajectory~(\ref{traj1}) into Eq.~(\ref{wightman1}), we can get
\begin{equation}\label{wightman2}
\begin{aligned}
W(\tau_D,\tau'_D)=&-\frac{1}{4\pi^2}\big[\frac{1}{(\gamma_D\Delta\tau-i\epsilon)^2-4R_D^2\sin^2(\gamma_D\omega_D\Delta\tau/2)}\\
&-\frac{1}{(\gamma_D\Delta\tau-i\epsilon)^2-4R_D^2\sin^2(\gamma_D\omega_D\Delta\tau/2)-4\Delta z^2}\big]\;,
\end{aligned}
\end{equation}
where $\Delta \tau=\tau _D-\tau '_D$.

Substituting the  Eq.~(\ref{wightman2}) into Eq.~(\ref{probty}), we get the expression for transition probability $P_D$ (see Appendix~{\ref{Derivation-PD}} for detail)
\begin{align}\label{PAPB}
\begin{aligned}
P_D=&K_D\int_0^{\infty }dx\frac{e^{-\alpha x^2}\cos (\beta x)(x^2-\sin^2 x)}{x^2 (x^2-v_D ^2 \sin^2 x)}\, +\frac{\lambda ^2 |\omega _{D}|\sigma}{4\pi ^{3/2} \gamma}{\rm{PV}} \int_0^{\infty }dx \frac{e^{-\alpha  x^2} \cos (\beta x)}{x^2-v_{D}^2\sin ^2(x)-\omega _{D}^2\Delta z^2 } \, \\
&+\frac{\lambda^2}{4\pi}\Big[e^{-\Omega^2\sigma^2}-\sqrt{\pi}\Omega\sigma {\rm{Erfc}}\big(\Omega\sigma\big)\Big]+\frac{\lambda ^2|\omega _{D}| \sigma}{4 \sqrt{\pi} \gamma}\frac{e^{-\alpha  S^2} \sin (\beta  S)}{2S-v_{D}^2 \sin (2S)},
\end{aligned}
\end{align}
where

\begin{equation}
\alpha=\frac{1}{\sigma^2\omega_D^2\gamma_D^2}=\frac{R_D}{a_D\sigma^2}\;,~~
\beta=\frac{2\Omega}{\gamma_D|\omega_D|}\;,~~K_D=\frac{\lambda^2v_D^2\gamma_D|\omega_D|\sigma}{4\pi^{3/2}}=\frac{\lambda^2v_Da_D\sigma}{4\pi^{3/2}\gamma_D}\;,
\end{equation}
and ${\rm{Erfc}}(x)=1-{\rm{Erf}}(x)$ is the complementary the error function with ${\rm{Erf}}(x):=\int_{0}^{x} 2e^{-t^2}dt/\sqrt{\pi}$. $S$ is the solution of the equation $x^2-v_{D}^2\sin^2x-\omega_D^2\Delta z^2=0$.

To make a cross-comparison, we consider the following uniformly accelerated detector trajectory~\cite{Birrell:1982,Crispino:2008,Rizzuto:2009,Salton-Man:2015}
\begin{equation}\label{un-traj}
x_D:=\{t=a_D^{-1}\sinh(a_D\tau)\;,~~x=a_D^{-1}\cosh(a_D\tau)\;,~~y={0}\;,~~z={\Delta z}\}\;,
\end{equation}
where $a_D$ still denotes the magnitude of the linear constant acceleration.
\begin{figure*}
\centering
\subfigure[$\Omega\sigma=0.10,~\Delta z/\sigma=0.20$]{\label{pvsromega01z02}
\includegraphics[width=0.30\textwidth]{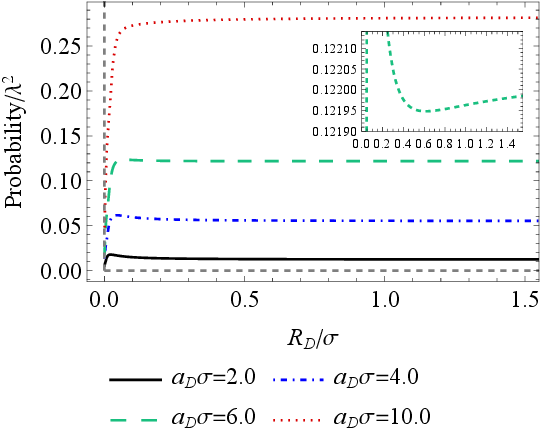}}\hspace{0.01\textwidth}
\subfigure[$\Omega\sigma=0.10,~\Delta z/\sigma=3.00$]{\label{pvsromega01z3}
\includegraphics[width=0.30\textwidth]{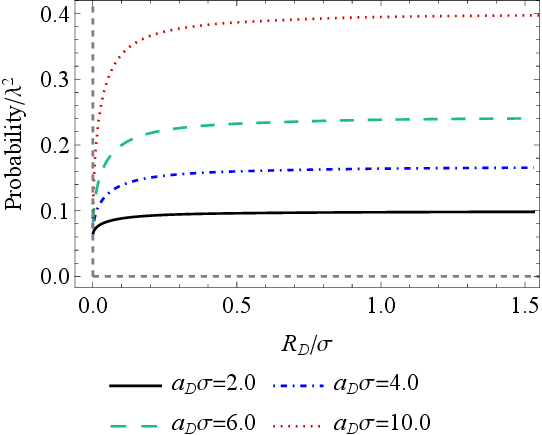}}\hspace{0.01\textwidth}
\subfigure[$\Omega\sigma=0.10,~\Delta z/\sigma=10.00$]{\label{pvsromega01z10}
\includegraphics[width=0.30\textwidth]{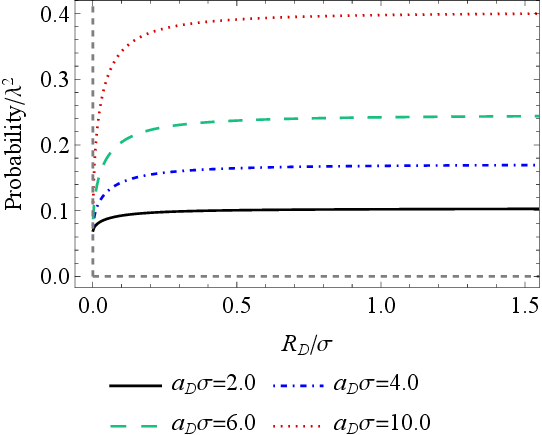}}\hspace{0.01\textwidth}
\subfigure[$\Omega\sigma=1.80,~\Delta z/\sigma=0.20$]{\label{pvsromega18z02}
\includegraphics[width=0.30\textwidth]{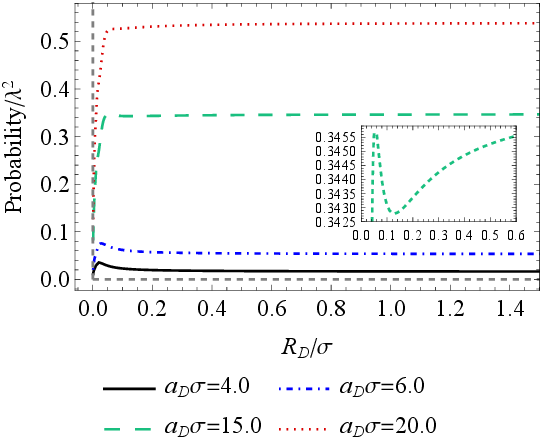}}\hspace{0.01\textwidth}
\subfigure[$\Omega\sigma=1.80,~\Delta z/\sigma=3.00$]{\label{pvsromega18z3}
\includegraphics[width=0.30\textwidth]{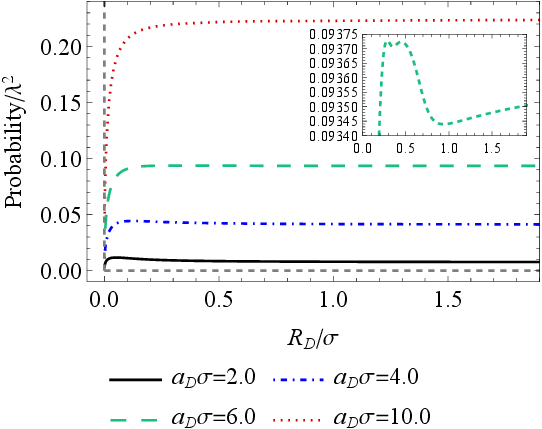}}\hspace{0.01\textwidth}
\subfigure[$\Omega\sigma=1.80,~\Delta z/\sigma=10.00$]{\label{pvsromega18z10}
\includegraphics[width=0.30\textwidth]{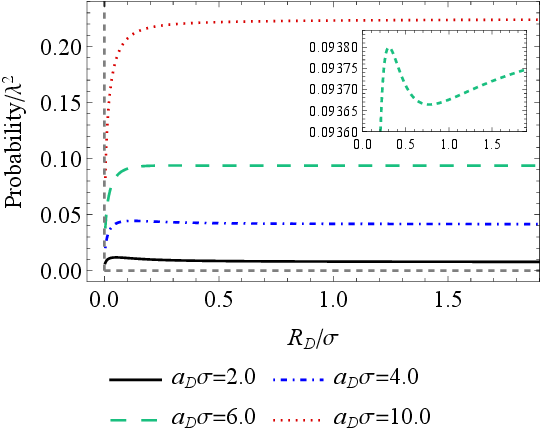}}
\caption{The transition probability of a UDW detector is plotted as a function of $R_D/\sigma$ for different values of $a_D\sigma$ with $\Omega\sigma=\{0.10, 1.80\}$ and $\Delta z/\sigma=\{0.20, 3.00, 10.00\}$. In each plot, the different colored curves correspond to different acceleration values. These behaviors can be characterized by certain critical values.}\label{pvsr}
\end{figure*}

In Fig.~\ref{pvsr}, we describe the transition probability as a function of $R_D/\sigma$ for different values of $a_D\sigma$ with fixed $\Omega\sigma=\{0.10, 1.80\}$ and $\Delta z/\sigma=\{0.20, 3.00, 10.00\}$. For small $\Omega\sigma=0.10$ and small $\Delta z/\sigma=0.2$ in Fig. \ref{pvsromega01z02}, as $R_D/\sigma$ increases, the transition probability increases and approaches to a stable nonzero value for large $a_D\sigma> a_{D_c}\sigma$ ($a_{D_c}\sigma\approx 6.931$). As $a_D\sigma$ decrease, the transition probability initially increases, then decreases, and finally increases to a stable value. With decreasing $a_D\sigma$, the transition probability first increases, then decreases to a stable value. There exists a single peak for not large $a_D\sigma$. However, it will increase to a stable value with a larger $\Delta z/\sigma$ for any nonzero $a_D\sigma$ as shown in Figs.~\ref{pvsromega01z3} and \ref{pvsromega01z10}, which is similar to the free space case \cite{ZJL:2020}. There is a competitive relationship between the boundary effect and the acceleration effect. When the acceleration is large, the behavior caused by the boundary is suppressed. Our results can reduce to the results of Ref. \cite{ZJL:2020} for far from the barrier. Compared with the Refs. \cite{ZJL:2020,LZH:2021}, we also study the case of larger energy gap. For a larger $\Omega\sigma=1.80$, the transition probability exhibits behavior similar to that in Fig. \ref{pvsromega01z02} for small $\Delta z/\sigma$. As $\Delta z/\sigma$ increases, there will exist more than one peak for some $a_D\sigma$ as shown in Fig. \ref{pvsromega18z3}. For a large $\Delta z/\sigma$, the feature is also similar to the case of Fig. \ref{pvsromega01z02}. The results of a larger energy gap are obviously different from the results for small energy gap \cite{ZJL:2020}. The peaks occur in two situations. Firstly, the vacuum fluctuations can be modified by a reflecting boundary and the acceleration of the detector, and the resulting distortions may cause the behavior. Secondly, the detector with larger energy gap $\Omega \sigma$ may be subject to greater uncertainty in the vacuum and thus exhibits the behavior. This behavior is reminiscent of that seen in Ref. \cite{Bozanic:2023}, where the peak is nonexistent for a small $\Omega\sigma$ but becomes manifest when $\Omega\sigma$ exceeds a certain value. The presence of the peak is contingent on larger values of $\Omega\sigma$ relative to $a_D\sigma$, and the peak disappears for a large $a_D\sigma$.

\begin{figure*}[!ht]
\centering
\subfigure[$\Omega\sigma=0.10,~\Delta z/\sigma=0.05$]{\label{pvsaomega01z005}
\includegraphics[width=0.3\textwidth]{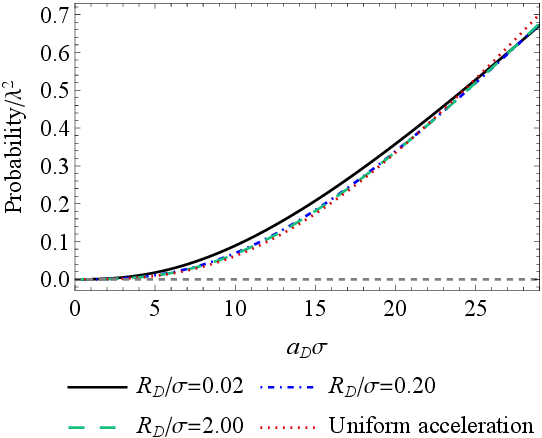}}\hspace{0.01\textwidth}
\subfigure[$\Omega\sigma=0.10,~\Delta z/\sigma=0.20$]{\label{pvsaomega01z02}
\includegraphics[width=0.3\textwidth]{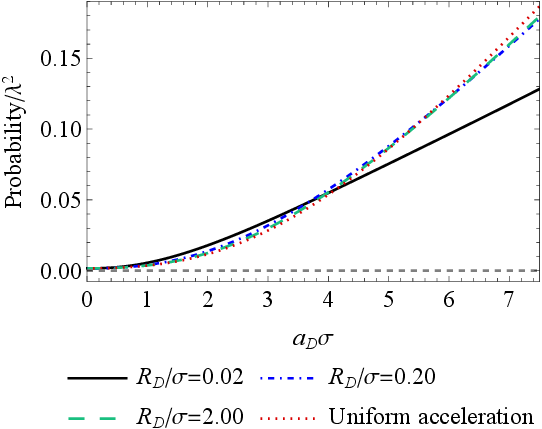}}\hspace{0.01\textwidth}
\subfigure[$\Omega\sigma=0.10,~\Delta z/\sigma=3.00$]{\label{pvsaomega01z3}
\includegraphics[width=0.3\textwidth]{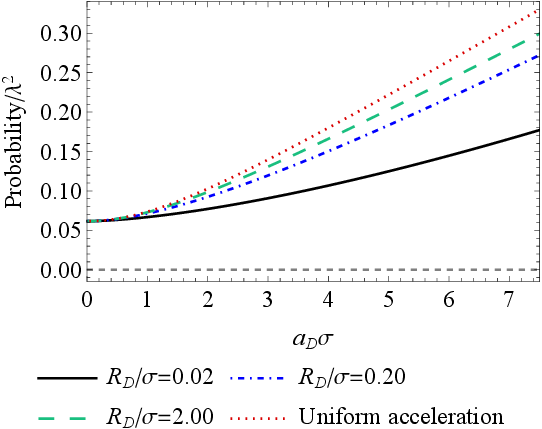}}\hspace{0.01\textwidth}
\subfigure[$\Omega\sigma=1.80,~\Delta z/\sigma=0.05$]{\label{pvsaomega18z005}
\includegraphics[width=0.3\textwidth]{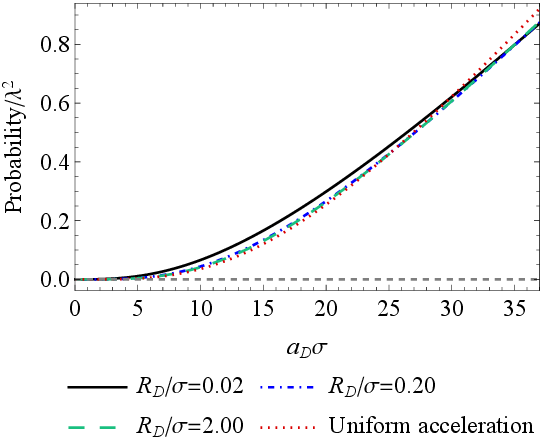}}\hspace{0.01\textwidth}
\subfigure[$\Omega\sigma=1.80,~\Delta z/\sigma=0.20$]{\label{pvsaomega18z02}
\includegraphics[width=0.3\textwidth]{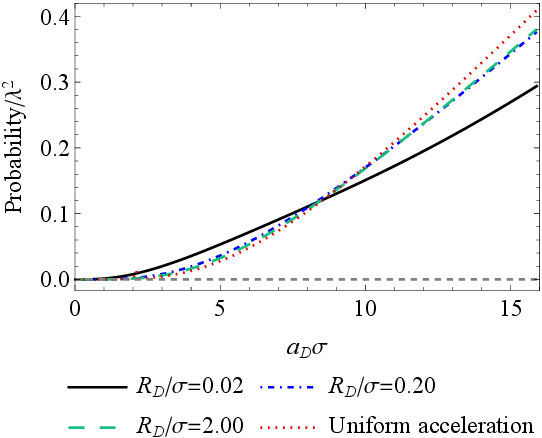}}\hspace{0.01\textwidth}
\subfigure[$\Omega\sigma=1.80,~\Delta z/\sigma=3.00$]{\label{pvsaomega18z3}
\includegraphics[width=0.3\textwidth]{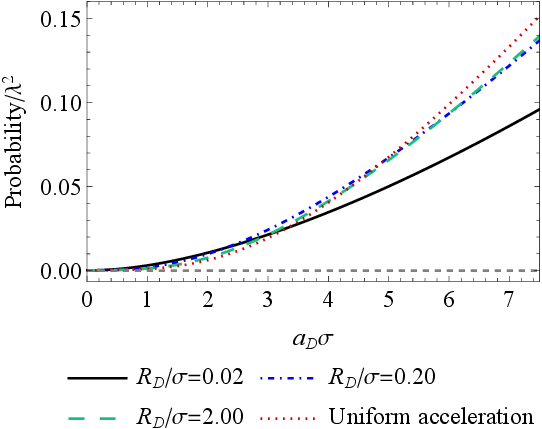}}
\caption{The transition probability of a UDW detector is plotted as a function of acceleration with $\Omega\sigma=\{0.10, 1.80\}$ for $R_D/\sigma=\{0.02, 0.20, 2.00\}$ and $\Delta z/\sigma=\{0.05, 0.20, 3.00\}$. The additional red line represents the case of uniform acceleration. The transition probability of a detector increases as the acceleration is increased.}\label{pvsa}
\end{figure*}

\begin{figure*}[!ht]
\centering
\subfigure[$\Delta{z}/\sigma=0.05$]{\label{pvsomegaa10z005}
\includegraphics[width=0.3\textwidth]{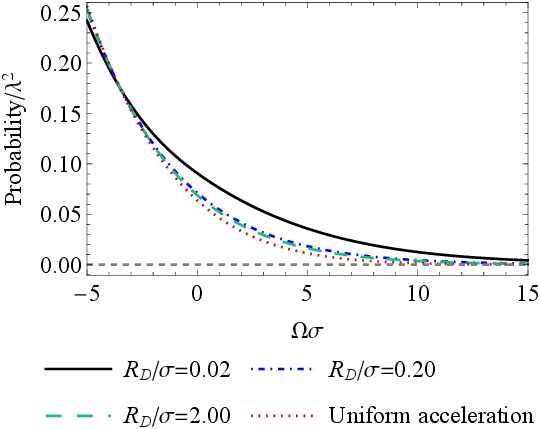}}\hspace{0.01\textwidth}
\subfigure[$\Delta{z}/\sigma=0.20$]{\label{pvsomegaa10z02}
\includegraphics[width=0.3\textwidth]{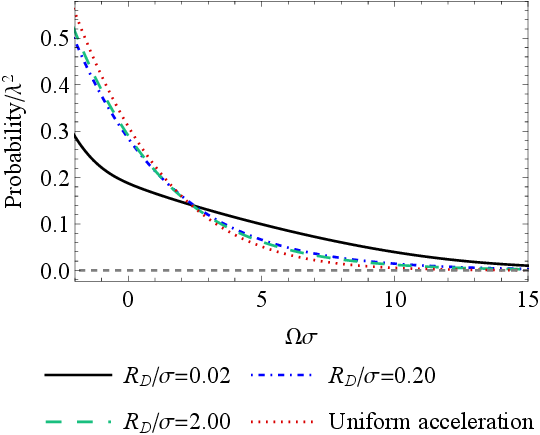}}\hspace{0.01\textwidth}
\subfigure[$\Delta{z}/\sigma=3.00$]{\label{pvsomegaa10z3}
\includegraphics[width=0.3\textwidth]{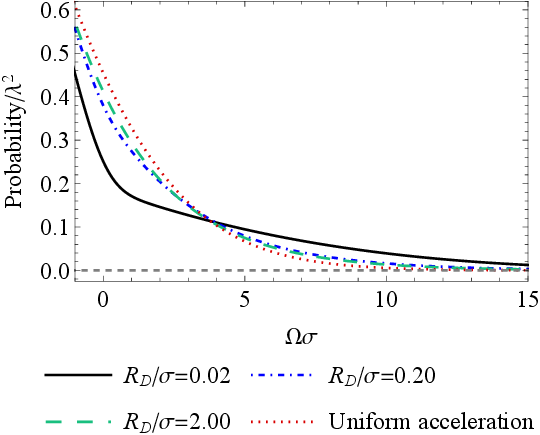}}
\caption{The transition probability of a UDW detector is plotted as a function of $\Omega\sigma$ with $a_{D}\sigma=10.0$ and $\Delta z/\sigma=\{0.05, 0.20, 3.00\}$ for both circularly and uniformly accelerated motion. The red line represents the uniformly accelerated case. The positive (negative) energy gaps correspond to that detector prepared in its ground (excited) state prior to interacting with the field. The transition probability decreases as $\Omega\sigma$ increases.}\label{pvsomega}
\end{figure*}
We plot the transition probability of a detector as a function of $a_{D}\sigma$ for different $\Delta z/\sigma$ and $\Omega\sigma$ in both the circular and linear motion scenarios, as shown in Fig. \ref{pvsa}. The transition probability of a detector is increasing function of acceleration. For a small $\Omega\sigma$ such as $\Omega\sigma=0.10$ in the top row, there exists a critical value $\Delta z_{c}/\sigma$ ($\Delta z_{c}/\sigma\approx 0.814$). When $\Delta z/\sigma<\Delta z_{c}/\sigma$, as $a_D\sigma$ increases, a larger transition probability can be obtained first for a smaller trajectory radius, which is opposite to the case of large $a_D\sigma$. There are intersections for different curves, which is different from unbounded case \cite{ZJL:2020}. As we amplify $\Delta z/\sigma$, the intersections are shifted to the left. When $\Delta z/\sigma>\Delta z_{c}/\sigma$, there is no intersection for all curves with nonzero $a_{D}\sigma$, which means a larger transition probability can be acquired for a larger trajectory radius with every nonzero $a_D\sigma$. However, with a larger $\Omega\sigma >\Omega_{c}\sigma $ ($\Omega_{c}\sigma\approx0.707$) such as $\Omega\sigma=1.80$ in the bottom row, the curves always exist intersections with different $R_D/\sigma$ for any $\Delta z/\sigma$, which is similar to Ref. \cite{Bozanic:2023} and different from the small energy gap in the Refs. \cite{ZJL:2020,LZH:2021}. The intersections in Fig. \ref{pvsa} correspond to the oscillatory behaviors shown in Fig.~\ref{pvsr}. For example, we can observe this feature from the oscillation in Fig. \ref{pvsromega01z02}. Before reaching a stable value, the same transition probability can correspond to different $R_D/\sigma$. For a very large $\Omega\sigma$, the transition probability goes to zero, which can be seen from Fig. \ref{pvsomega}. We depict the transition probability of a detector as a function of $\Omega\sigma$ in Fig. \ref{pvsomega}. The transition probability decreases with the increase of $\Omega\sigma$. As $\Delta z/\sigma$ increases, the intersections are shifted to the right until not move.
\begin{figure*}
\centering
\subfigure[$\Delta z/\sigma=0.70,~\Omega\sigma=0.10$]{\label{dpvsaomega01z07}
\includegraphics[width=0.35\textwidth]{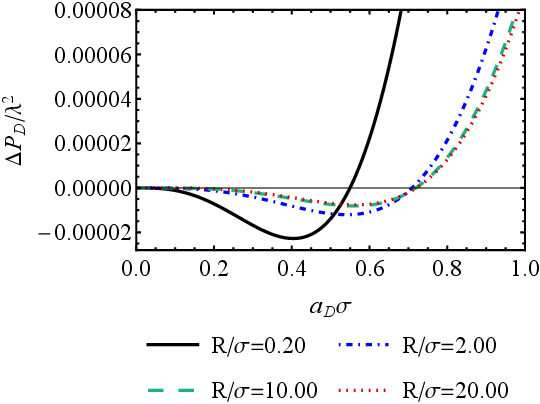}}\hspace{0.01\textwidth}
\subfigure[$a\sigma=1.00,~R/\sigma=2.00$]{\label{dpvsdzr2a1}
\includegraphics[width=0.35\textwidth]{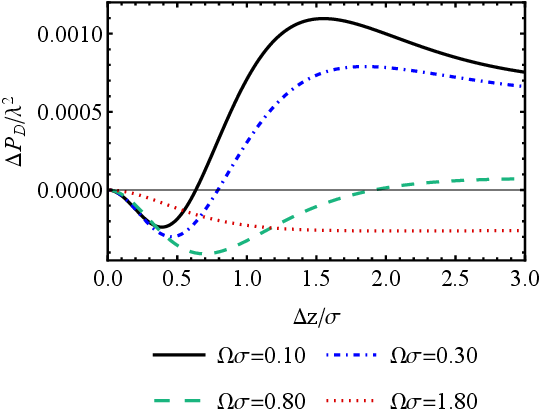}}\hspace{0.01\textwidth}
\caption{The difference value of the transition probability as a function of $a_{D}\sigma$ and $\Delta z/\sigma$, respectively. $\Delta P_D$ represents the result of the uniformly accelerated case minus the circular motion case.}\label{dp}
\end{figure*}

As shown in Fig. \ref{pvsa}, the transition probabilities of circular motion and linear uniformly acceleration cases intersect at certain points, which are the optimal circumstances for the circular case to simulate the uniformly accelerating results. For example, at $a\sigma=21.136$, the transition probability for linear uniformly acceleration equals that of the circular motion with $R_D/\sigma=2.00$, $\Omega\sigma=0.10$ and $\Delta z/\sigma=0.05$. Furthermore, we display the difference value of the transition probability in circular motion and the linear uniformly accelerating case in Fig. \ref{dp}. The difference value is the result of the uniformly accelerated case minus that of the circular motion case. By fixing certain parameters, we can choose a suitable region where circular motion can simulate the results of uniform acceleration.

\section{Entanglement harvesting of UDW detectors in the circular motion near the reflecting boundary}

In this section, we study the entanglement harvesting of two circularly accelerated detectors with a boundary. For simplicity, we mainly focus on the spacetime trajectories of the detectors in coaxial rotation (see Fig.~\ref{orbit}).
	
We assume that detectors $A$ and $B$ with angular velocities $\omega_A$ and $\omega_B$ rotate around the common axis with the radii $R_A$ and $R_B$. The spacetime trajectories of such two detectors can be parameterized by their proper time $\tau_A$ and $\tau_B$
\begin{align}\label{traj2}
&x_A:=\{t=\tau_A\gamma_A\;~,~x=R_A\cos(\omega_A\tau_A\gamma_A)\;~,~
	y=R_A\sin(\omega_A\tau_A\gamma_A)\;~,~z=\Delta z\}\;,\nonumber\\
&x_B:=\{t=\tau_B\gamma_B\;~,~x=R_B\cos(\omega_B\tau_B\gamma_B)\;~,~
	y=R_B\sin(\omega_B\tau_B\gamma_B)\;~,~z=\Delta z+\Delta{d}\}\;,
\end{align}
where $\gamma_A$ and $\gamma_B$ are corresponding Lorentz factors of detectors $A$ and $B$, and $\Delta{d}$ is the distance between two detectors.
	
We can get the transition probabilities of two detectors by using Eq.~(\ref{PAPB}) and the spacetime trajectories (\ref{traj2}). Substituting the trajectories Eq.~(\ref{traj2}) and the Wightman function Eq.~(\ref{wightman1}) into Eq.~(\ref{defX}), we get the nonlocal correlation term $X$. After some algebraic manipulations, $X$ can be written as
\begin{align}\label{Xint1-1}
X&=-\frac{\lambda^2\sigma^2}{4\pi^2\gamma_A\gamma_B}\int_{-\infty}^{\infty} d\tilde{u} \int_{0}^{\infty} d\tilde{s}\bigg\{\exp\Big[\frac{-\gamma_A^2\tilde{u}^2-\gamma_B^2(\tilde{s}-\tilde{u})^2}{2\gamma_A^2\gamma_B^2}\Big]\exp\Big[\frac{i(\tilde{s}-\tilde{u})\sigma\Omega}{\gamma_A}-\frac{i\tilde{u}\sigma\Omega}{\gamma_B}\Big]\nonumber\\
&\times(\frac{1}{f_{AB}(\tilde{u},\tilde{s})}-\frac{1}{f_{AB}(\tilde{u},\tilde{s})+4\Delta{d}\Delta{z}+4\Delta{z}^2})+\exp\Big[\frac{-\gamma_B^2\tilde{u}^2-\gamma_A^2(\tilde{s}-\tilde{u})^2}{2\gamma_A^2\gamma_B^2}\Big] \nonumber\\
&\times\exp\Big[\frac{i(\tilde{s}-\tilde{u})\sigma\Omega}{\gamma_B}-\frac{i\tilde{u}\sigma\Omega}{\gamma_A}\Big](\frac{1}{f_{BA}(\tilde{u},\tilde{s})}-\frac{1}{f_{BA}(\tilde{u},\tilde{s})+4\Delta d\Delta z+4\Delta z^2})\bigg\}\;,
\end{align}
where the auxiliary functions are
\begin{equation}
f_{AB}(\tilde{u},\tilde{s})=\Delta d^2+R_A^2+R_B^2-2R_AR_B\cos\big({\tilde{u}\omega_A\sigma}-{\tilde{u}\omega_B\sigma-\tilde{s}\omega_A\sigma}\big)-\sigma^2(\tilde{s}+i\epsilon)^2\;,
\end{equation}
\begin{equation}
f_{BA}(\tilde{u},\tilde{s})=\Delta d^2+R_A^2+R_B^2-2R_AR_B\cos\big({\tilde{u}\omega_A\sigma}-{\tilde{u}\omega_B\sigma+\tilde{s}\omega_B\sigma}\big)-\sigma^2(\tilde{s}+i\epsilon)^2\;.
\end{equation}
		
We suppose that the two detectors are completely synchronously rotating around $z$-axis, i.e., $\omega_A=\omega_B=\omega$, then Eq.~(\ref{Xint1-1}) can be written as
\begin{align}\label{Xint1-2}
X=&-\frac{\lambda^2\sigma^2}{\pi^{3/2}\sqrt{2(\gamma_A^2+\gamma_B^2)}}\exp\Big[\frac{-\sigma^2\Omega^2(\gamma_A+\gamma_B)^2}{2(\gamma_A^2+\gamma_B^2)}\Big]\nonumber\\
&\int_{0}^{\infty}d\tilde{s}\,\cos\Big[\frac{\tilde{s}\sigma\Omega(\gamma_A-\gamma_B)}{\gamma_A^2+\gamma_B^2}\Big]\times\exp\Big[\frac{-\tilde{s}^2}{2(\gamma_A^2+\gamma_B^2)}\Big](\frac{1}{f(\tilde{u},\tilde{s})}-\frac{1}{f(\tilde{u},\tilde{s})+4\Delta d\Delta z+4\Delta z^2})\;,
\end{align}
where
\begin{align}
	f(\tilde{u},\tilde{s})=\Delta d^2+R_A^2+R_B^2-2R_AR_B\cos\big(\tilde{s}\omega\sigma\big)-\sigma^2(\tilde{s}+i\epsilon)^2\;.
\end{align}

It is hard to obtain analytical results for Eq.~(\ref{Xint1-1}) and Eq.~(\ref{Xint1-2}). Therefore, we need numerical evaluations. We can obtain the concurrence from Eq.~(\ref{con1}) by evaluating the values of $P_{D}$ and $X$ (see Appendix~\ref{Derivation-PD}).

We study the entanglement harvesting phenomenon in the situation where two detectors are rotating with the same acceleration and trajectory radius, i.e, $a_A=a_B=a\;, R_A=R_B=R\;.$ For simplicity, we consider that the two detectors are completely comoving ($\omega_A=\omega_B=\omega$), therefore we obtain
\begin{equation}\label{Xint1-4}
X\big|_{a,R}=- \frac{\lambda^2\sigma^2e^{-\sigma^2\Omega^2}}{2\pi^{3/2}\gamma} \int_{0}^{\infty}d\tilde{s}e^{-\tilde{s}^2/(4\gamma^2)}(\frac{1}{f(\tilde{s})\big|_{a,R}}-\frac{1}{f(\tilde{s})\big|_{a,R}+4\Delta d\Delta z+4\Delta z^2})\;,
\end{equation}
where
\begin{equation}
f(\tilde{s})\big|_{a,R}=\Delta d^2+4R^2\sin^2(\tilde{s}\omega\sigma/2)-\sigma^2(\tilde{s}+i\epsilon)^2\;.
\end{equation}

For comparison with the case of linear uniformly accelerated motion, we consider the following trajectory for uniform acceleration
\begin{align}\label{traj-ua}
&x_A:=\{t=a^{-1}\sinh(a\tau_A)\;,~x=a^{-1}\cosh(a\tau_A)\;,~y=0\;,~z=\Delta z\}\;,
		\nonumber\\
&x_B:=\{t=a^{-1}\sinh(a\tau_B)\;,~x=a^{-1}\cosh(a\tau_B)\;,~y=0\;,~z=\Delta z+\Delta{d}\}\;,
\end{align}
where  $a$ still represents the magnitude of acceleration and $\Delta{d}$ denotes the separation between two detectors.

\begin{figure*}[!ht]
\centering
\subfigure[$\Delta{d}/\sigma=0.20,\Delta{z}/\sigma=0.20$]{\label{cvsro01d02z02}
\includegraphics[width=0.35\textwidth]{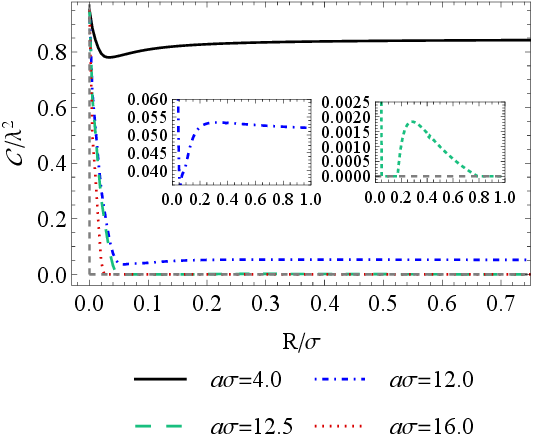}}\hspace{0.01\textwidth}
\subfigure[$\Delta{d}/\sigma=0.20,\Delta{z}/\sigma=3.00$]{\label{cvsro01d02z3}
\includegraphics[width=0.35\textwidth]{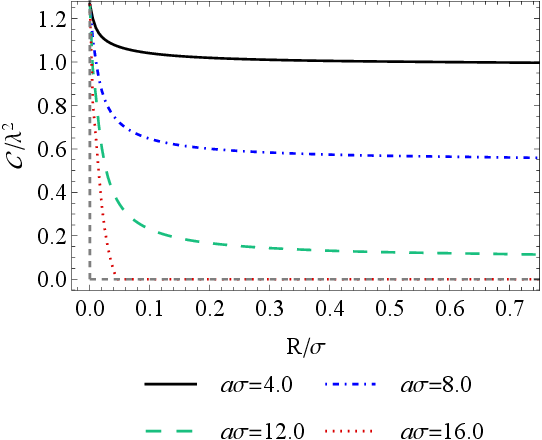}}\hspace{0.01\textwidth}
\subfigure[$\Delta{d}/\sigma=1.00,\Delta{z}/\sigma=0.20$]{\label{cvsro01d1z02}
\includegraphics[width=0.35\textwidth]{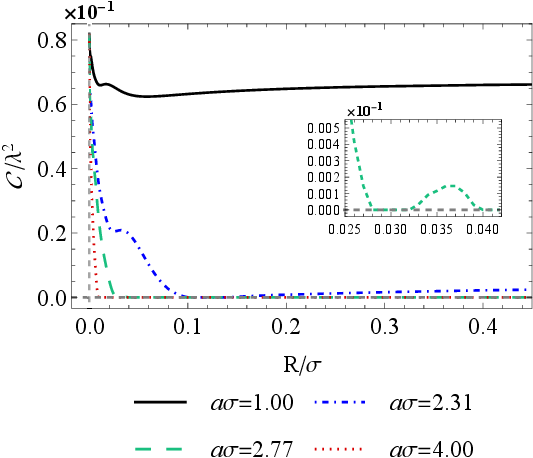}}\hspace{0.01\textwidth}
\subfigure[$\Delta{d}/\sigma=1.00,\Delta{z}/\sigma=3.00$]{\label{cvsro01d1z3}
\includegraphics[width=0.35\textwidth]{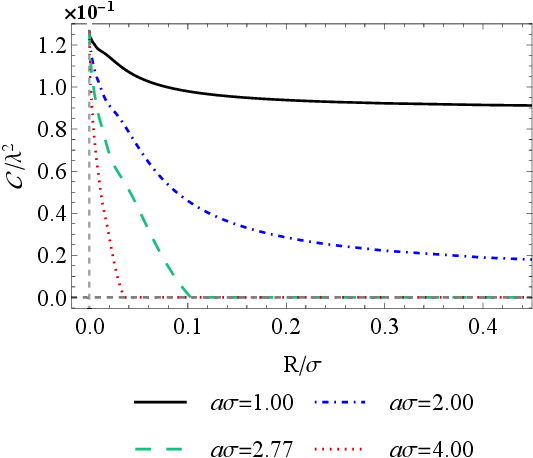}}
\caption{The concurrence ${\cal{C}}(\rho_{AB})/\lambda^2$ is plotted as a function of $R/\sigma$ for different values of $a\sigma$ with $\Delta{d}/\sigma=\{0.20, 1.00\}$ and $\Delta{z}/\sigma=\{0.20, 3.00\}$. Here, we have set $\Omega\sigma=0.10$. In each plot, the different colored curves correspond to different acceleration values. We observe that the concurrence rapidly decreases to zero for large $a\sigma$.}\label{cvsr}
\end{figure*}

\begin{figure*}[!ht]
\centering
\subfigure[$\Delta{z}/\sigma=0.20$]{\label{cvsro18d02z02}
\includegraphics[width=0.35\textwidth]{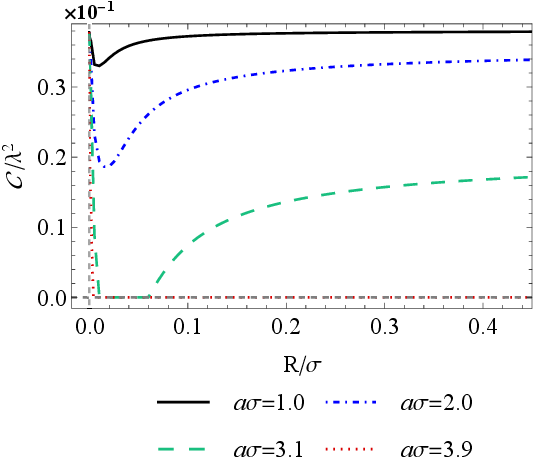}}\hspace{0.01\textwidth}
\subfigure[$\Delta{z}/\sigma=3.00$]{\label{cvsro18d02z3}
\includegraphics[width=0.35\textwidth]{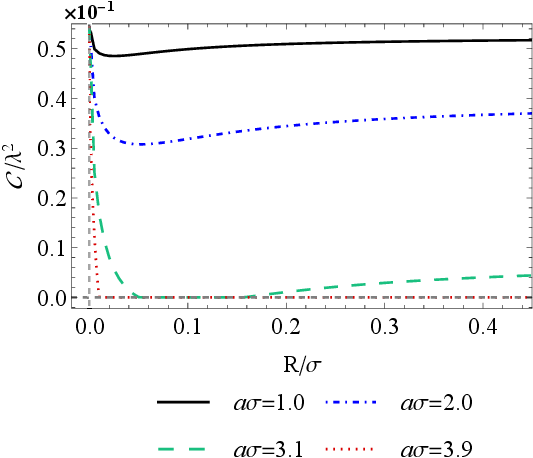}}
\caption{The concurrence ${\cal{C}}(\rho_{AB})/\lambda^2$ is plotted as a function of $R/\sigma$ for different values of $a\sigma$ with $\Omega\sigma=1.80$, $\Delta{d}/\sigma=0.20$, and $\Delta{z}/\sigma=\{0.20, 3.00\}$. The concurrence rapidly decreases to zero for large $a\sigma$.}\label{cvsr2}
\end{figure*}

\begin{figure*}[!htbp]
\centering
\subfigure[$a\sigma=1.00,\Delta{d}/\sigma=0.20$]{\label{cvszo01d02a1}
\includegraphics[width=0.35\textwidth]{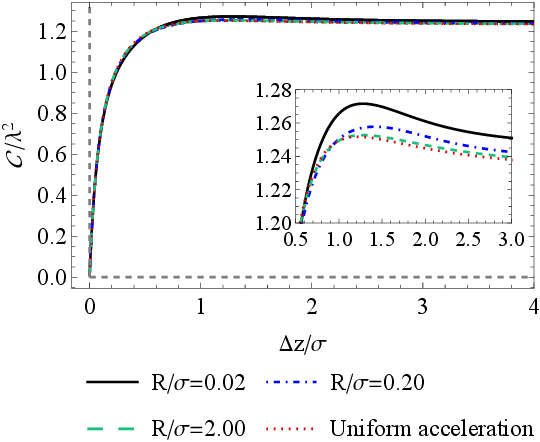}}\hspace{0.01\textwidth}
\subfigure[$a\sigma=1.00,\Delta{d}/\sigma=1.05$]{\label{cvszo01d105a1}
\includegraphics[width=0.35\textwidth]{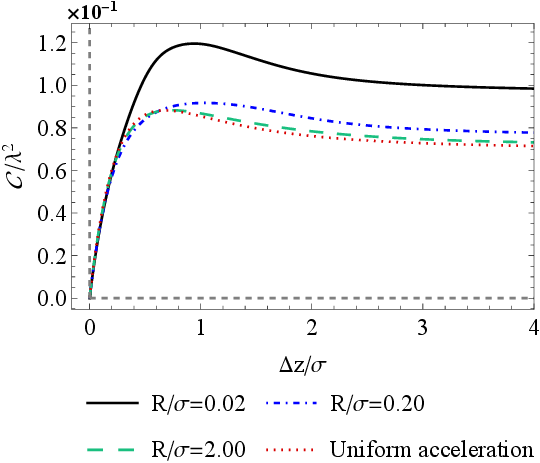}}\hspace{0.01\textwidth}
\subfigure[$a\sigma=2.00,\Delta{d}/\sigma=0.20$]{\label{cvszo01d02a2}
\includegraphics[width=0.35\textwidth]{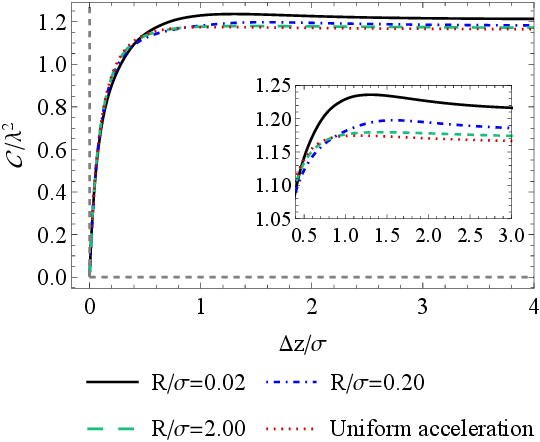}}\hspace{0.01\textwidth}
\subfigure[$a\sigma=2.00,\Delta{d}/\sigma=1.05$]{\label{cvszo01d105a2}
\includegraphics[width=0.35\textwidth]{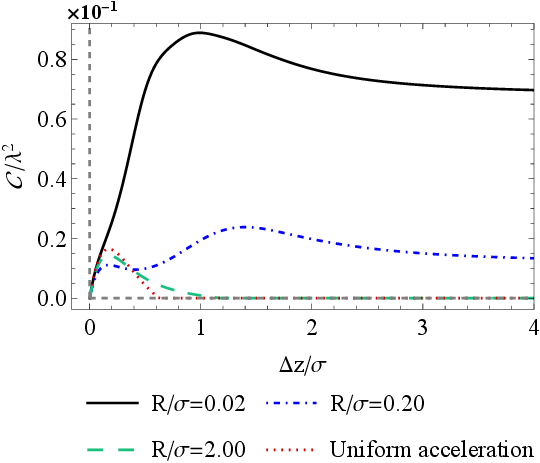}}
\caption{The concurrence ${\cal{C}}(\rho_{AB})/\lambda^2$ is plotted as a function of $\Delta{z}/\sigma$ with different values of $a\sigma$ and $\Delta{d}/\sigma$. We have set $\Omega\sigma=0.10$ for both circularly and uniformly accelerated motion. The red line describes the situation of uniform acceleration. We observe that peaks occur under certain conditions.}\label{cvsz}
\end{figure*}

\begin{figure*}[!ht]
\centering
\subfigure[$a\sigma=1.00, \Delta{d}/\sigma=0.01$]{\label{cvszo25a1d001}
\includegraphics[width=0.35\textwidth]{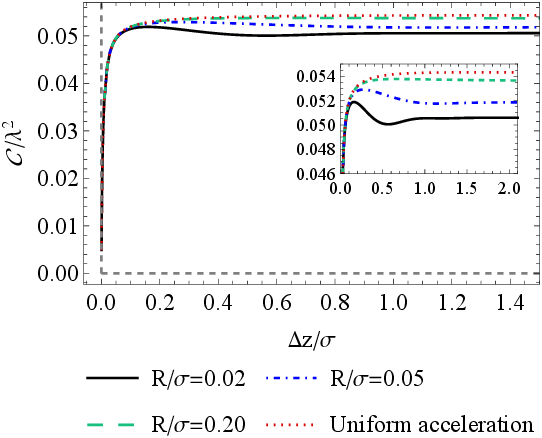}}\hspace{0.01\textwidth}
\subfigure[$a\sigma=1.00, \Delta{d}/\sigma=0.15$]{\label{cvszo25a1d015}
\includegraphics[width=0.35\textwidth]{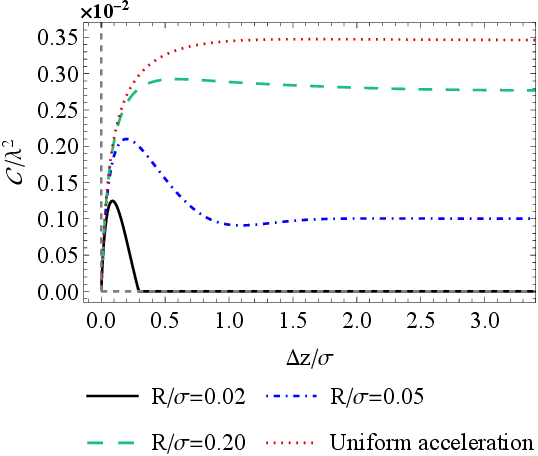}}\hspace{0.01\textwidth}
\subfigure[$a\sigma=2.00, \Delta{d}/\sigma=0.01$]{\label{cvszo25a2d001}
\includegraphics[width=0.35\textwidth]{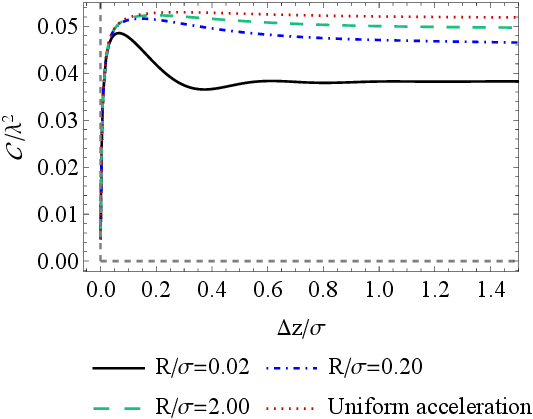}}\hspace{0.01\textwidth}
\subfigure[$a\sigma=2.00, \Delta{d}/\sigma=0.15$]{\label{cvszo25a2d015}
\includegraphics[width=0.35\textwidth]{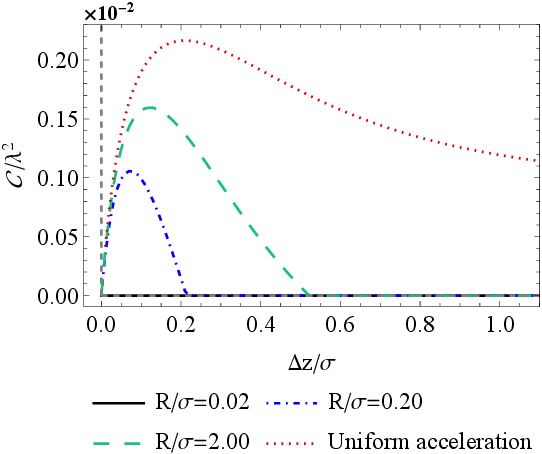}}
\caption{The concurrence ${\cal{C}}(\rho_{AB})/\lambda^2$ is plotted as a function of $\Delta{z}/\sigma$ for different values of $a\sigma$ and $\Delta{d}/\sigma$. We have set $\Omega\sigma=2.50$ for both circularly and uniformly accelerated motion. The red line describes the uniformly accelerated situation. The oscillatory behaviors may occur near the boundary.}\label{cvsz2}
\end{figure*}

\begin{figure*}[!ht]
\centering
\subfigure[$\Delta{d}/\sigma=0.50, \Delta{z}/\sigma=0.20$]{\label{cvsomegaa1d05z02}
\includegraphics[width=0.35\textwidth]{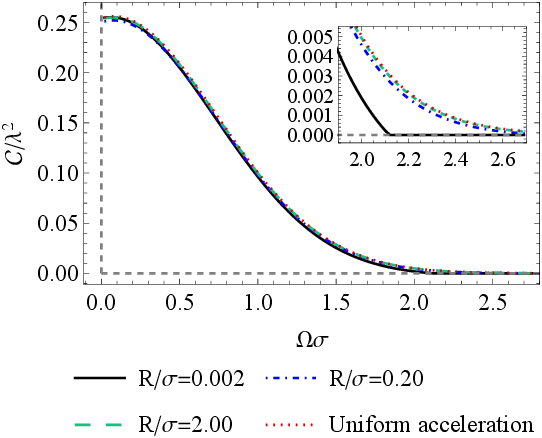}}\hspace{0.01\textwidth}
\subfigure[$\Delta{d}/\sigma=0.50, \Delta{z}/\sigma=3.00$]{\label{cvsomegaa1d05z3}
\includegraphics[width=0.35\textwidth]{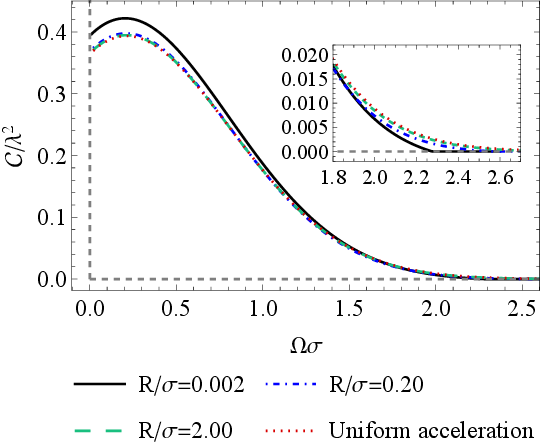}}\hspace{0.01\textwidth}
\subfigure[$\Delta{d}/\sigma=1.50, \Delta{z}/\sigma=0.20$]{\label{cvsomegaa1d15z02}
\includegraphics[width=0.35\textwidth]{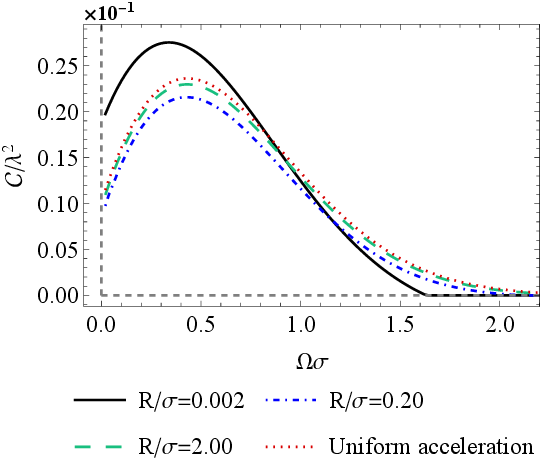}}\hspace{0.01\textwidth}
\subfigure[$\Delta{d}/\sigma=1.50, \Delta{z}/\sigma=3.00$]{\label{cvsomegaa1d15z3}
\includegraphics[width=0.35\textwidth]{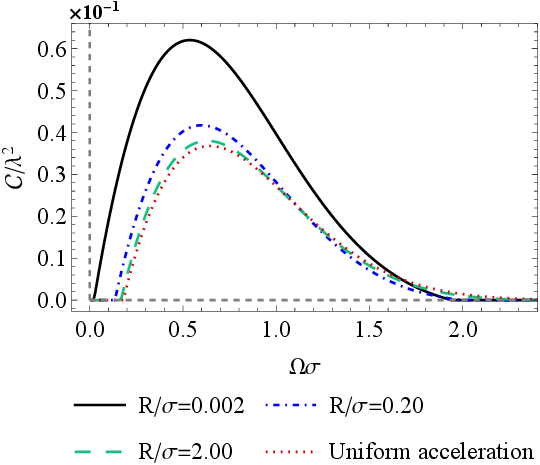}}\hspace{0.01\textwidth}
\subfigure[$\Delta{d}/\sigma=2.00, \Delta{z}/\sigma=0.20$]{\label{cvsomegaa1d2z02}
\includegraphics[width=0.35\textwidth]{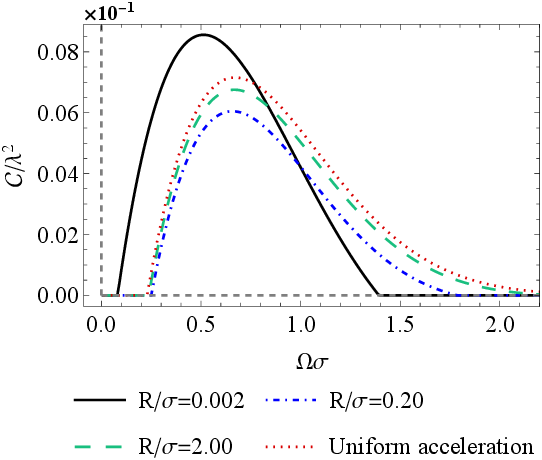}}\hspace{0.01\textwidth}
\subfigure[$\Delta{d}/\sigma=2.00, \Delta{z}/\sigma=3.00$]{\label{cvsomegaa1d2z3}
\includegraphics[width=0.35\textwidth]{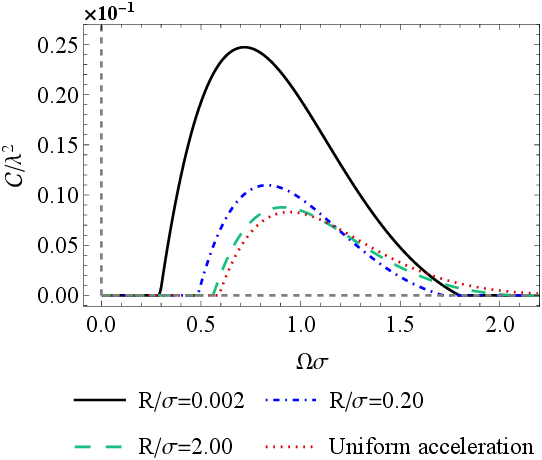}}
\caption{The concurrence ${\cal{C}}(\rho_{AB})/\lambda^2$ is plotted as a function of $\Omega\sigma$ for $\Delta{d}/\sigma=\{0.50, 1.50, 2.00\}$ and $\Delta z/\sigma=\{0.20, 3.00\}$. We have set $a\sigma=1.00$ for both circularly and uniformly accelerated motion. There is a peak of concurrence for each curve. For a larger $\Delta{d}/\sigma=2.00$, the concurrence with a small $\Omega\sigma$ takes zero for different $\Delta{z}/\sigma$.}\label{cvsomega}
\end{figure*}

We depict the concurrence as a function of $R/\sigma$ for a small $\Omega\sigma=0.10$ in Fig. \ref{cvsr} and a larger $\Omega\sigma=1.80$ in Fig. \ref{cvsr2}. For small $\Delta{d}/\sigma=0.20$ and small $\Delta{z}/\sigma=0.20$ in Fig. \ref{cvsro01d02z02}, the entanglement (concurrence) rapidly decreases to zero as $R/\sigma$ increases under very large $a\sigma$. As we decrease $a\sigma$, the concurrence first decreases, then increases, but finally reduces to a stable value. Decreasing $a\sigma$, the concurrence first decreases, then increases to a stable value. For larger $\Delta{d}/\sigma=1.00$ and small $\Delta{z}/\sigma=0.20$ as shown in \ref{cvsro01d1z02}, with a not large $a\sigma$, the concurrence first decreases, then increases, subsequently decreases again, and finally increases to a stable value. The feature for large $a\sigma$ is similar to the case of small $\Delta{d}/\sigma$ and small $\Delta{z}/\sigma$. From Figs. \ref{cvsro01d02z02} and \ref{cvsro01d1z02}, we observe that the concurrence has oscillatory behavior, which is different from the free space \cite{ZJL:2020}. In Figs. \ref{cvsro01d02z3} and \ref{cvsro01d1z3}, for larger $\Delta{z}/\sigma=3.00$, increasing $R/\sigma$, the concurrence decreases and finally arrives at a stable value, which is similar to the case without any boundary \cite{ZJL:2020}. For a larger $\Omega\sigma$ such as $\Omega\sigma=1.80$ in Fig. \ref{cvsr2}, there exist similar characteristics for any nonzero $\Delta{z}/\sigma$. For a very large $a\sigma$, the concurrence will decrease to zero. As $a\sigma$ decrease, the concurrence first decreases, then increases, and finally reaches a stable value. This result is different from the small energy gap situation, which also displays in Ref. \cite{ZJL:2020}. It should be noted that the concurrence in some situations first decreases to zero, then maintains zero, and finally increases to a stable value. When the acceleration effect is dominant, the curve monotonically decreases. The behavior caused by the boundary is suppressed.

\begin{figure*}[!ht]
\centering
\subfigure[$\Omega\sigma=0.10, \Delta z/\sigma=0.20$]{\label{cvsao01d02z02}
\includegraphics[width=0.35\textwidth]{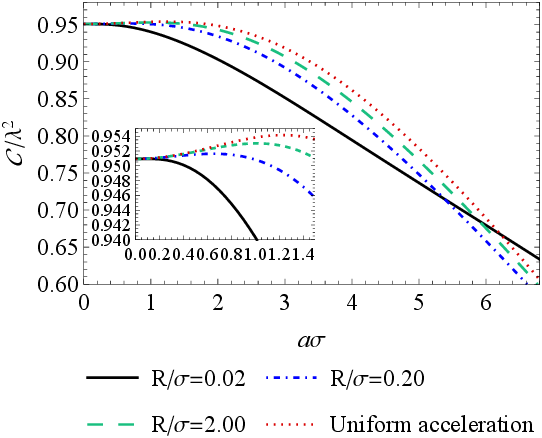}}\hspace{0.02\textwidth}
\subfigure[$\Omega\sigma=0.10, \Delta z/\sigma=3.00$]{\label{cvsao01d02z3}
\includegraphics[width=0.35\textwidth]{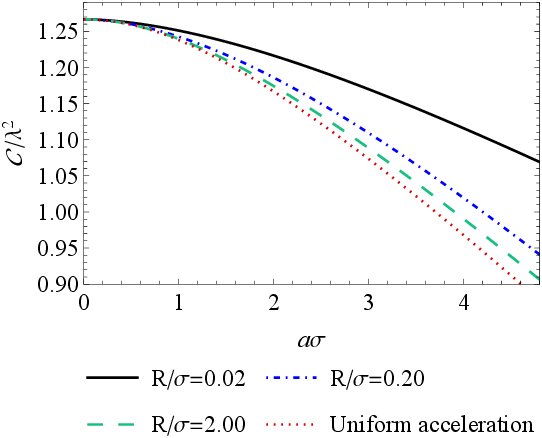}}\hspace{0.02\textwidth}
\subfigure[$\Omega\sigma=1.80, \Delta z/\sigma=0.20$]{\label{cvsao18d02z02}
\includegraphics[width=0.35\textwidth]{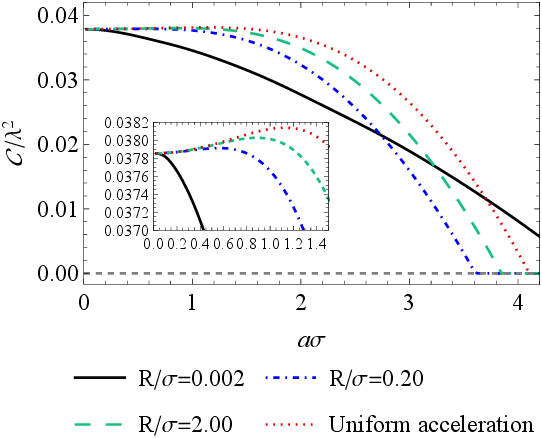}}\hspace{0.02\textwidth}
\subfigure[$\Omega\sigma=1.80, \Delta z/\sigma=3.00$]{\label{cvsao18d02z3}
\includegraphics[width=0.35\textwidth]{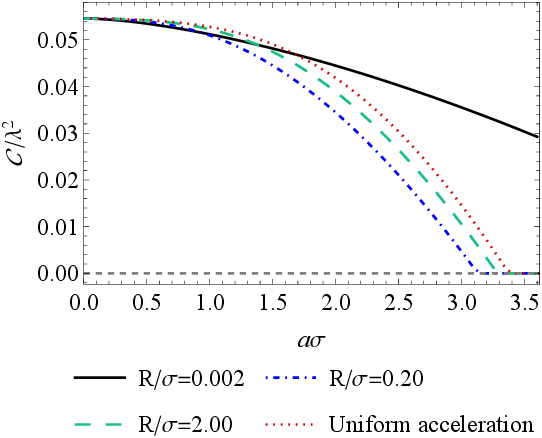}}
\caption{The concurrence ${\cal{C}}(\rho_{AB})/\lambda^2$ is plotted as a function of $a\sigma$ for $\Omega\sigma=\{0.10, 1.80\}$ and $\Delta{z}/\sigma=\{0.20, 3.00\}$. We have set $\Delta{d}/\sigma=0.20$ for both circularly and uniformly accelerated motion. There exists a peak value for small value of $\Delta{z}/\sigma$. There is no intersection with nonzero $a\sigma$ for small $\Omega\sigma$ and larger $\Delta{z}/\sigma$.}\label{cvsa}
\end{figure*}

\begin{figure*}
\centering
\subfigure[$\Delta z/\sigma=0.20,~\Omega\sigma=0.10$]{\label{dcvsad02z02omega01}
\includegraphics[width=0.35\textwidth]{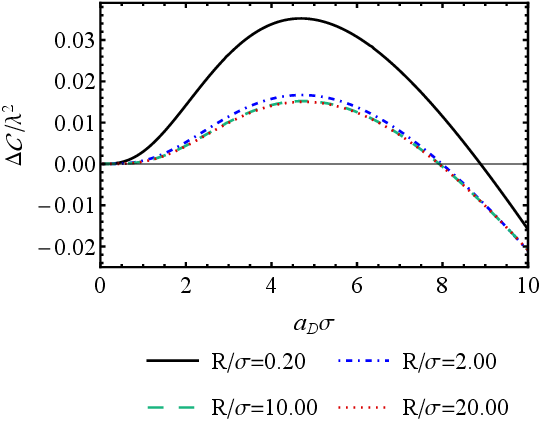}}\hspace{0.01\textwidth}
\subfigure[$a\sigma=1.00,~R/\sigma=2.00$]{\label{dcvszr2a1d02}
\includegraphics[width=0.35\textwidth]{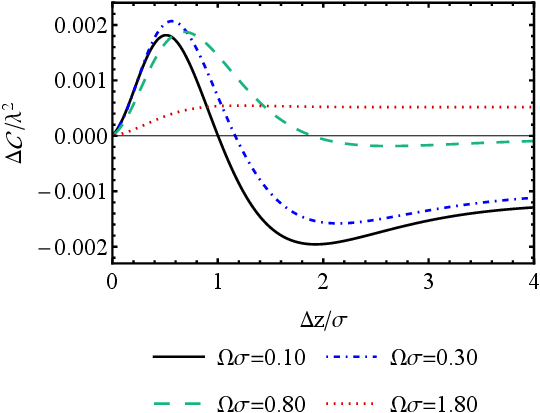}}\hspace{0.01\textwidth}
\caption{The difference value of the concurrence is plotted as a function of $a_{D}\sigma$ and $\Delta z/\sigma$, respectively. Here, $\Delta C$ is the result of the uniformly accelerated case minus the circular motion case.}\label{dc}
\end{figure*}

In Fig. \ref{cvsz}, we display the concurrence as a function of $\Delta{z}/\sigma$ for a small $\Omega\sigma=0.10$. For small $\Delta{d}/\sigma=0.20$ or small $a\sigma=1$, the concurrence first increases, then decreases, and finally reaches a stable value. This behavior is reminiscent of that seen in Ref. \cite{CW:2019}, where the entanglement harvested between two detectors can be greater in the presence of a mirror as compared to free space and there may exist a peak in the concurrence at a certain distance from the mirror. When we take a larger $\Delta{d}/\sigma=1.05$ and a not small $a\sigma=2.00$, there may exist two peaks for certain values of $R/\sigma$, which is different from Ref. \cite{LZH:2021}. In Fig. \ref{cvsz2}, we describe the concurrence as a function of $\Delta{z}/\sigma$ for a larger $\Omega\sigma=2.50$. There may exist more than one peak for a very small $\Delta{d}/\sigma=0.01$ or small $a\sigma=1$. For not very small $\Delta{d}/\sigma$ and not small $a\sigma$, there only exists no more than one peak. From Figs. \ref{cvsz} and \ref{cvsz2}, we observe that oscillatory behaviors emerge near the boundary, which caused by the vacuum fluctuations of the field and states of the detectors.

We depict the concurrence as a function of $\Omega\sigma$ for $a\sigma=1.00$ in Fig. \ref{cvsomega}. For a large $\Omega\sigma$, the concurrence goes to zero with any $\Delta{d}/\sigma$ and $\Delta z/\sigma$. For $\Delta{d}/\sigma=0.50$, the concurrence first increases from a nonzero value, then decreases to zero. As we amplify $\Delta{d}/\sigma=1.50$, we observe that the concurrence with a small $\Omega\sigma$ may take zero for a larger $\Delta z/\sigma$ from Fig. \ref{cvsomegaa1d15z3}. When we take a larger $\Delta{d}/\sigma=2.00$, the concurrence with a small $\Omega\sigma$ takes zero for any nonzero $\Delta z/\sigma$.

In Fig. \ref{cvsa}, we describe the concurrence as a function of $a\sigma$ with $\Delta{d}/\sigma=0.20$. For a small $\Delta{z}/\sigma=0.20$, the concurrence first increases, and then decreases to zero. There exist peaks which are caused by the vacuum fluctuations and states of the detectors. This is different from unbounded case in Ref. \cite{ZJL:2020}. For a larger $\Delta{z}/\sigma=3.00$, as $a\sigma$ increases, the concurrence decreases to zero. There is no intersection with nonzero $a\sigma$ for small $\Omega\sigma$ and larger $\Delta{z}/\sigma$ as shown in Fig. \ref{cvsao01d02z3}. As $\Delta{d}/\sigma$ increases, the concurrence goes to zero.

\begin{table}[htbp]\footnotesize
  \centering
    \begin{tabular}{|p{6em}|p{19em}|p{19em}|}
    \hline
    \multicolumn{1}{|l|}{Parameters} & \multicolumn{1}{l|}{Transition probability } & \multicolumn{1}{l|}{Concurrence} \bigstrut\\
    \hline
    Small $a \sigma$  & Increases non-monotonically with $R/\sigma$ for small $\Delta{z}/\sigma$ and small $\Omega \sigma$ before reaching stable value    & Decreases non-monotonically with $R/\sigma$ for small $\Delta{z}/\sigma$ and small $\Omega \sigma$ before reaching stable value \bigstrut\\
    \hline
    Large $\Omega \sigma$ & Increases non-monotonically with $R/\sigma$ for small $a \sigma$ and any
nonzero $\Delta{z}/\sigma$ before reaching stable value & Decreases non-monotonically with $R/\sigma$ for small $a \sigma$ and  any
nonzero $\Delta{z}/\sigma$ before reaching stable value \bigstrut\\
    \hline
    Small $\Delta{z}/\sigma$ & Increases monotonically with $a \sigma$ and exists
intersections for different $R/\sigma$   & May decrease non-monotonically with $a \sigma$ and exist
intersections for different $R/\sigma$ \bigstrut\\
    \hline
    \end{tabular} \label{tab}
      \caption{Main results of transition probability and concurrence.}
\end{table}

The concurrence of the detectors for circular motion case and linear uniformly acceleration case has intersections, which are the optimal circumstances for the circular motion to simulate the uniformly accelerating results. In addition, we depict the difference value of the concurrence for the circular motion and the linear uniformly accelerating case in Fig. \ref{dc}. We can use the circular motion to simulate the uniformly accelerating results by choosing a suitable region. We summarize the main results in Table I.

\section{Conclusion}

In this paper, we have studied the properties of the transition probability of a circularly accelerated UDW detector coupled with massless scalar fields with a reflecting boundary. As an inevitable result by the uncertainty principle, vacuum fluctuates and the modifications may have rich structures. The presence of a boundary in flat spacetime modifies the spacetime topology and cause changes of vacuum fluctuations. For a small $\Omega\sigma$ and a small $\Delta z/\sigma$, with the increase of $R_D/\sigma$, the transition probability increases directly to a stable value for $a_D\sigma>a_{D_c}\sigma$. When we decrease $a_D\sigma$, the transition probability first increases, then decreases, and finally increases to a stable value. Decreasing $a_D\sigma$, the transition probability first increases, then decreases to a stable value. The vacuum fluctuations can be modified by a reflecting boundary and the acceleration of the detector, and the resulting distortions may cause the peaks. It will increase to a stable value with a larger $\Delta z/\sigma$ for any nonzero $a_D\sigma$. For a larger $\Omega\sigma$, there will exist more than one peak for some $a_D\sigma$ and $\Delta z/\sigma$. The detector with larger energy gap $\Omega \sigma$ may be subject to greater uncertainty in the vacuum and thus exhibits peaks. Furthermore, for a small $\Omega\sigma$, there exists a critical value $\Delta z_{c}/\sigma$. When $\Delta z/\sigma<\Delta z_{c}/\sigma$, as we amplify $a_D\sigma$, a larger transition probability can be achieved first at a smaller trajectory radius, which is opposite to the case of large $a_D\sigma$. As $\Delta z/\sigma$ increases, the intersections are shifted to the left. When $\Delta z/\sigma>\Delta z_{c}/\sigma$, a larger transition probability can be obtained for a larger trajectory radius with every nonzero $a_D\sigma$. However, with a larger $\Omega\sigma >\Omega_{c}\sigma $, the curves always exist intersections with different $R_D/\sigma$ for any nonzero $\Delta z/\sigma$. The relation between the transition probability and $\Omega\sigma$ has been considered. For a very large $\Omega\sigma$, the transition probability goes to zero. As $\Delta z/\sigma$ increases, the intersections are shifted to the right until not move.

We also have investigated the entanglement harvesting phenomenon of two detectors with a boundary. We have discussed that the two detectors are rotating around a common axis with the same acceleration, trajectory radius and angular velocity. The relation between entanglement harvesting and $R/\sigma$ has been taken into account. We first considered a small $\Omega\sigma$. For larger $\Delta{z}/\sigma$, the entanglement harvesting decreases and finally arrives at a stable value. For small $\Delta{d}/\sigma$ and small $\Delta{z}/\sigma$, as $R/\sigma$ increases, the entanglement harvesting in a very large $a\sigma$ rapidly decays to zero. When $a\sigma$ decreases, the entanglement harvesting first decreases, then increases, but finally decreases to a stable value. As we decrease $a\sigma$, the entanglement harvesting first decreases, then increases, and finally goes to a stable value. For large $\Delta{d}/\sigma$ and small $\Delta{z}/\sigma$, with a not large $a\sigma$, the entanglement harvesting first decreases, then increases, then decreases, and finally increases to a stable value. The feature of large $a\sigma$ is similar to that of small $\Delta{d}/\sigma$ and small $\Delta{z}/\sigma$. We also considered a larger $\Omega\sigma$. For nonzero $\Delta{z}/\sigma$, the entanglement harvesting will decrease to zero with a large $a\sigma$. As we decrease $a\sigma$, the entanglement harvesting first decreases, then increases, and finally reaches a stable value. The influence of $\Delta z/\sigma$ has been investigated. Due to the presence of a boundary in flat spacetime, the modes of the field as a result of the superposition of the propagating incident and reflected modes, the features observed properties of the detector can produce rich structures. We first discussed a small $\Omega\sigma$. For small $\Delta{d}/\sigma$ or small $a\sigma$, the entanglement harvesting first increases, then decrease, and finally arrives at a stable value. For not small $a\sigma$ and larger $\Delta{d}/\sigma$, the entanglement harvesting may exist two peaks for some $R/\sigma$. For a larger $\Omega\sigma$, there may exist more than one peak with small $a\sigma$ or very small $\Delta{d}/\sigma$. The entanglement harvesting is no more than one peak for not small $a\sigma$ and not very small $\Delta{d}/\sigma$. The influence of $\Omega\sigma$ for fixed $a\sigma$ has been studied. For a large $\Omega\sigma$, the entanglement harvesting goes to zero. As $\Delta{d}/\sigma$ increases, the entanglement harvesting with a larger $\Delta z/\sigma$ may take zero at a small $\Omega\sigma$. For a larger $\Delta{d}/\sigma$, the entanglement harvesting with a small $\Omega\sigma$ takes zero for any nonzero $\Delta z/\sigma$. We have discussed the effect of $a\sigma$ for not large $\Delta{d}/\sigma$. For a small $\Delta{z}/\sigma$, the entanglement harvesting first increases, and then decreases to zero. There exists a peak value. For a larger $\Delta{z}/\sigma$, as $a\sigma$ increases, the entanglement harvesting decreases to zero. There is no intersection with a nonzero $a\sigma$ for small $\Omega\sigma$ and larger $\Delta{z}/\sigma$. As $\Delta{d}/\sigma$ increases, the entanglement harvesting goes to zero. The features observed properties of the detectors are closely related to the vacuum fluctuations of the field and states of the detectors. Although the Unruh effect has theoretical significance, it has yet to be verified experimentally. The main challenge hindering its experimental verification lies in the large acceleration required to produce experimentally measurable temperatures. The circular trajectory has attracted considerable attention for potential experimental realizations of the Unruh effect. Under the proper conditions, the response and concurrence of detectors in circular motion very closely approximates the linear uniformly accelerating results. Choosing a suitable region, we can use the circular motion to simulate the uniformly accelerating results. In the context of an experiment that transfers nonclassical correlations from an ultra-cold atom system to a pair of pulsed laser probes, Gooding et al. described how the entanglement harvesting protocol can be realized \cite{Gooding}. Lindel et al. showed that genuine and communication based entanglement harvesting from the vacuum field is possible with state-of-the-art electro-optic sampling experimental setups \cite{Lindel}. The investigation presented in this paper can help us better understand the relativistic and quantum effects.

\begin{acknowledgments}

This work was supported by the Nature Science Foundation of Shaanxi Province, China under Grant No. 2023-JC-YB-016 and the National Natural Science Foundation of China under Grant No. 11705144.

\end{acknowledgments}

\appendix
\section{Derivation of  $P_D$ and $X$}\label{Derivation-PD}
In this appendix, we derive $P_D$ and $X$ from Eq.~(\ref{probty}) and Eq.~(\ref{defX}) respectively.
\subsection{The transition probability $P_D$}\label{Derivation-PD1}
Setting $u=\tau_{D}$ and $s=\tau_{D}-\tau_{D}'$, we have
\begin{align}\label{PA-A1}
P_D&=\lambda^2\int_{-\infty}^{\infty}{du}\chi_D(u)\int_{-\infty}^{\infty}{ds}\chi_D(u-s)e^{-i\Omega{s}}W(s)\nonumber\\
&=\lambda^2\sqrt{\pi}\sigma\int_{-\infty}^{\infty}{ds}e^{-i\Omega{s}}e^{-s^2/(4\sigma^2)}W(s)\;.
\end{align}
Inserting Eq. (\ref{wightman2}) into Eq.~(\ref{PA-A1}) and using $x=\gamma_{D}|\omega_D|s/2$, the transition probability can be written as $P_D=P_1+P_2$. The first term read
\begin{align}\label{PA-A11}
	 P_1=\frac{\lambda^2\sigma|\omega_D|}{8\pi^{3/2}\gamma_D}\int_{-\infty}^{\infty}dx\frac{{e}^{-ix\beta}e^{-x^2\alpha}}{v_{D}^2\sin^2x-(x-i\epsilon)^2}\;,
\end{align}
and the second term can be written as
\begin{align}\label{PA-A21}
	 P_2=\frac{\lambda^2\sigma|\omega_D|}{8\pi^{3/2}\gamma_D}\int_{-\infty}^{\infty}dx\frac{{e}^{-ix\beta}e^{-x^2\alpha}}{(x-i\epsilon)^2-v_{D}^2\sin^2x-\omega_D^2\Delta z^2}\;,
\end{align}
where $\alpha:=1/(\sigma^2\omega_D^2\gamma_D^2)$, $\beta:={2\Omega}/(\gamma_D|\omega_D|)$.
$P_1$ can be rewritten as
\begin{align}\label{PA-A12}
	P_1&=\frac{\lambda^2\sigma|\omega_D|}{8\pi^{3/2}\gamma_D}\int_{-\infty}^{\infty}dx\bigg[\frac{{e}^{-i x\beta}e^{-x^2\alpha}}{v_D^2\sin^2x-(x-i\epsilon)^2}+\frac{{e}^{-i x\beta}e^{-x^2\alpha}}{(1-v_D^2)(x-i\epsilon)^2}-\frac{{e}^{-i x\beta}e^{-x^2\alpha}}{(1-v_D^2)(x-i\epsilon)^2}\bigg]\;
	\nonumber\\
	 &=K_D\int_{0}^{\infty}dx\frac{\cos(x\beta)e^{-x^2\alpha}(x^2-\sin^2x)}{x^2(x^2-v_D^2\sin^2x)}-\frac{\lambda^2\sigma|\omega_D|}{8\pi^{3/2}\gamma_D}\int_{-\infty}^{\infty}dx\frac{{e}^{-ix\beta}e^{-x^2\alpha}}{(1-v_D^2)(x-i\epsilon)^2},
\end{align}
where $K_D:={\lambda^2v_D^2\gamma_D|\omega_D|\sigma}/(4\pi^{3/2})$.
In second line of Eq. (\ref{PA-A12}), we have neglected $i\epsilon$ since the integral is now regular. The second term can be expressed as
\begin{align}\label{PA-A13}
    &-\frac{\lambda^2\sigma|\omega_D|}{8\pi^{3/2}\gamma_D}\int_{-\infty}^{\infty}dx\frac{{e}^{-ix\beta}e^{-x^2\alpha}}{(1-v_D^2)(x-i\epsilon)^2}
	\nonumber\\
	 &=-\frac{\lambda^2\sigma\gamma_D|\omega_D|}{8\pi^{3/2}}\int_{-\infty}^{\infty}dx\frac{{e}^{-ix\beta}e^{-x^2\alpha}}{x^2}+\frac{i\lambda^2\sigma\gamma_D|\omega_D|}{8\pi^{1/2}}\int_{-\infty}^{\infty}dx{e}^{-ix\beta}e^{-x^2\alpha}\delta^{(1)}(x).\
\end{align}
We have used the following identity,
\begin{equation}\label{id0}
	\frac{1}{(x\pm{i}\epsilon)^n}=\frac{1}{x^n}\pm\frac{(-1)^n}{(n-1)!}{i\pi}\delta^{(n-1)}(x).
\end{equation}
Considering the definition of a distribution $g$ acting on a test function $f$
\begin{equation}
	\langle{g},{f}\rangle:=\int_{-\infty}^{\infty}g(x)f(x)dx\;,
\end{equation}
we have the following identities for a distribution function~\cite{MMST:2016,Bogolubov:1990}
\begin{equation}\label{id1}
	\big\langle{\frac{1}{x}},{f(x)}\big\rangle={\rm{PV}}\int_{-\infty}^{\infty}\frac{f(x)}{x}dx\;,
\end{equation}
\begin{equation}\label{id2}
	\big\langle{\frac{1}{x^2}},{f(x)}\big\rangle=\int_{0}^{\infty}dx\frac{f(x)+f(-x)-2f(0)}{x^2}\;,
\end{equation}
and
\begin{equation}\label{id3}
    \big\langle{\delta^{(n)}(x)},{f(x)}\big\rangle=(-1)^nf^{(n)}(0)\;,
\end{equation}
where $\rm{PV}$ is the principle value of an integral.
Therefore, by using Eq.~(\ref{id2}) and Eq.~(\ref{id3}),  Eq.~(\ref{PA-A13}) can be rewritten as
\begin{align}\label{PA-A14}
	 &-\frac{\lambda^2\sigma|\omega_D|}{8\pi^{3/2}\gamma_D}\int_{-\infty}^{\infty}dx\frac{{e}^{-ix\beta}e^{-x^2\alpha}}{(1-v_D^2)(x-i\epsilon)^2}=\frac{\lambda^2}{4\pi}\Big[e^{-\Omega^2\sigma^2}-\sqrt{\pi}\Omega\sigma\;\rm{Erfc}\big(\Omega\sigma\big)\Big]\;.
\end{align}
$P_2$ can be expressed as
\begin{align}\label{PA-A22}
	 P_2&=\frac{\lambda^2\sigma|\omega_D|}{8\pi^{3/2}\gamma_D}\big[\int_{-\infty}^{0}dx\frac{{e}^{-ix\beta}e^{-x^2\alpha}}{x^2-v_{D}^2\sin^2x-\omega_D^2\Delta z^2+i\epsilon}\ +\int_{0}^{\infty}dx\frac{{e}^{-ix\beta}e^{-x^2\alpha}}{x^2-v_{D}^2\sin^2x-\omega_D^2\Delta z^2-i\epsilon}\big]
	\nonumber\\
	&=\frac{\lambda^2\sigma|\omega_D|}{4\pi^{3/2}\gamma_D}{\rm{PV}} \int_{0}^{\infty}dx\frac{\cos(x\beta)e^{-x^2\alpha}}{x^2-v_{D}^2\sin^2x-\omega_D^2\Delta z^2}+\frac{\lambda^2\sigma|\omega_D|}{4\sqrt{\pi}\gamma_D}\frac{e^{-S^2\alpha}\sin(S\beta)}{2S-v_{D}^2\sin 2S}\;,
\end{align}
where $S$ is the solution of $x^2-v_{D}^2\sin^2x-\omega_D^2\Delta z^2=0$. We can obtain the expression of the transition probability given in Eq.~(\ref{PAPB}).
\subsection{The expression of $X$}\label{Derivation-PD2}
From the definition of $X$ in Eq.~(\ref{defX}), we have
\begin{align}
	X=&-\frac{\lambda^2}{\gamma_A\gamma_B}\int_{-\infty}^{\infty}dt\int_{-\infty}^{t}dt'
	\bigg[\chi_B(\tau_B(t)) \chi_A(\tau_A(t'))e^{-i(\Omega {t}/\gamma_B+\Omega{t'}/\gamma_A)} W\!\left(x_A(t'), x_B(t)\right) \nonumber \\
	&\quad + \chi_A(\tau_A(t)) \chi_B(\tau_B(t'))  e^{-i( \Omega{t}/\gamma_A  +\Omega{t'}/\gamma_B)} W\!\left(x_B(t'),x_A(t) \right)  \bigg]\nonumber\\
	=&-\frac{\lambda^2\sigma^2}{\gamma_A\gamma_B}\int_{-\infty}^{\infty}  d\tilde{u} \int_{0}^{\infty} d\tilde{s} \,\bigg[e^{-\tilde{u}^2(\gamma_B^{-2}+\gamma_A^{-2})/2}e^{-\tilde{s}^2/2\gamma_A^2}
	e^{\tilde{s}\tilde{u}/\gamma_A^2}e^{-i\tilde{u}\Omega\sigma[\gamma_B^{-1}+\gamma_A^{-1}]}e^{i\tilde{s}\Omega\sigma/\gamma_A}W\!\left(x_A( t' ), x_B(t )\right)\nonumber\\
	 &\quad+e^{-\tilde{u}^2(\gamma_A^{-2}+\gamma_B^{-2})/2}e^{-\tilde{s}^2/2\gamma_B^2}e^{\tilde{s}\tilde{u}/\gamma_B^2}e^{-i\tilde{u}\Omega\sigma[\gamma_A^{-1}+\gamma_B^{-1}]}e^{i\tilde{s}\Omega\sigma/\gamma_B}W\!\left(x_B( t' ), x_A(t )\right) \bigg]\;,\label{defX-A}
\end{align}
where we have used $\tilde{u}=t/\sigma,\tilde{s}=(t-t')/\sigma\;$. In particular, if the Wightman function is only dependent on $\tilde{s}$, Eq.~(\ref{defX-A}) can be further expressed as
\begin{align}
	X=&-
	\frac{\sqrt{2\pi}\lambda^2\sigma^2}{\sqrt{\gamma_A^2+\gamma_B^2}}
	\exp\Big[\frac{-\sigma^2\Omega^2(\gamma_A+\gamma_B)^2}{2\gamma_A^2+2\gamma_B^2}\Big]\int_{0}^{\infty}
	d\tilde{s} \bigg\{ \exp\Big[\frac{i\tilde{s}\sigma\Omega(\gamma_A
	-\gamma_B)}{\gamma_A^2+\gamma_B^2}\Big]\exp\Big[\frac{-\tilde{s}^2}{2(\gamma_A^2+\gamma_B^2)}\Big]\nonumber\\
    &\times{W}\!\left(x_A( t' ), x_B(t)\right)+\exp\Big[\frac{i\tilde{s}\sigma\Omega(\gamma_B
	-\gamma_A)}{\gamma_A^2+\gamma_B^2}\Big]\exp\Big[\frac{-\tilde{s}^2}{2(\gamma_A^2+\gamma_B^2)}\Big]W\!\left(x_B(t' ), x_A(t )\right) \bigg\}\;.
\end{align}
Using the explicit expression of the Wightman function, one can obtain  Eq.~(\ref{Xint1-2}) and Eq.~(\ref{Xint1-4}).


\end{document}